\documentclass[preprint,10pt,a4paper]{aastex}
\usepackage[colorlinks,urlcolor=blue,citecolor=blue,linkcolor=blue]{hyperref}
\usepackage{amsmath}
\usepackage{graphicx}
\usepackage{multicol}
\newcommand{\be}{\begin{equation}} 
\newcommand{\ee}{\end{equation}}

\newcommand{\ba}{\begin{eqnarray}}
\newcommand{\ea}{\end{eqnarray}}

\newcommand\eg{\textit{e.g.,\ }}

\newcommand{\Bf}{{magnetic field}}
\newcommand{\Bfs}{{magnetic fields}}
\newcommand{\Ef}{{electric  field}}

\newcommand{\NS}{neutron star}
\newcommand{\NSs}{{neutron stars}}
\newcommand{\EM}{electromagnetic}
\newcommand{\BH}{{black hole}}
\newcommand{\BHs}{{black holes}}
\newcommand{\Sc}{Schwarzschild}

\begin{document}

\title{Fermi GBM signal contemporaneous with GW150914 - an unlikely association}
\author{Maxim Lyutikov}
\affil{Department of Physics and Astronomy, Purdue University, 
 525 Northwestern Avenue,
West Lafayette, IN
47907-2036, USA; lyutikov@purdue.edu
}

\begin{abstract} 
The physical constraints required by the association of the Fermi GBM signal contemporaneous with GW150914 - radiative  power of $10^{49} $ erg s$^{-1}$, and corresponding   magnetic fields on the  black hole of the order of $10^{12}$ Gauss - are astrophysically highly implausible. Combined with the relatively high random probability of coincidence  of  0.22  percents,  we conclude that the \EM\ signal   is likely  unrelated to the BH merger.
\end{abstract}

\section{Electromagnetic signals accompanying merger of compact objects}

The report of the possible Fermi GBM  signal \citep{2016arXiv160203920C} associated with the merger of two \BHs\  detected by LIGO \citep[GW150914,][]{2016PhRvL.116f1102A} could be a ground-breaking discovery in high energy astrophysics. The claimed event lasts approximately a second, produces non-thermal emission in the keV-MeV range with, most importantly, overall luminosity of $L_{EM}\sim 10^{49}$ erg s$^{-1}$. The event was not detected by INTEGRAL  \citep {2016arXiv160204180S}, which was covering the  GW150914  region at the time of the GW trigger  and has larger  effective area above 100 keV.

The  luminosity implied by the  Fermi GBM detection exceeds by nearly 10 orders of magnitude the Eddington luminosity for a $M=60 M_\odot$ star. Thus, the \EM\ signal cannot be powered by a quasi-spherical accretion. An alternative possibility is that the accreting matter brings in the \Bf\ that extracts rotational and/or translational  energy of the central object(s). This process of \EM\ extraction of mechanical energy, first proposed by \cite{Blandford:1977} (see also \cite{2002luml.conf..381B}) has many advantages in generating clean, highly relativistic outflow.

If the  \EM\ signal from GW150914 is real, it is then necessary that the emitting plasma is highly relativistic \citep[by analogy with  Gamma Ray Bursts (GRBs), \eg][]{2004RvMP...76.1143P}. 
In our  case  the high compactness parameter at the source,
\be
l_c = {L_{EM} \sigma_T  \over 2 c G m_e M} = 1.5 \times 10^{13},
\ee
would require that the emission region propagates with the bulk Lorentz factor $\Gamma \approx 100$ - hence, a requirement of a clean, relativistic outflow \citep[][discussed the \EM\ model of 
 GRBs]{lb03,LyutikovJPh}. Below we concentrate on 
this  more realistic scenario of the \EM\ energy extraction from \BH(s);  pressure and neutrino-driven outflows \citep[\eg][]{2007ApJ...657..383C} would  fare even worse at the expected  low accretion rates.

The \EM\ counterparts of  mergers of compacts objects have been studied mostly for NS-NS binaries. These  include the suggested association with  short GRBs \citep[\eg][]{1992ApJ...395L..83N}, the precursor emission \cite[\eg][]{2001MNRAS.322..695H}, and the  long lasting post-merger emission originating  either  from a supermassive magnetar \citep{2011MNRAS.413.2031M} or a \BH\ that keeps its magnetic ``hair'' - the magnetic flux   \citep[][]{2011PhRvD..84h4019L,2013ApJ...768...63L,2014ApJ...788..186N}.
In the case of BH-BH merger, the  expected \EM\ counterparts can be produced through accretion via Blandford-Znajek mechanism \citep[][rotation of a \BH\ in externally-supplied \Bf]{Blandford:1977}, or through linear motion of  \BHs\ in \Bf\ \citep{2010Sci...329..927P,2011PhRvD..83f4001L}, or   a combination thereof \citep{2014PhRvD..89j4030M}. 

In all   cases mentioned above, the \EM\ power comes from the kinetic energy of the rotation  or the  linear motion  of the central source, converted into Poynting flux with the ``help'' of the \Bf\ (and subsequent dissipation and particle acceleration). Qualitatively,  the \EM\ power of relativistic outflows can be estimated as  \citep{Blandford:1977,2002luml.conf..381B} 
\be 
L_{EM} \approx V^2 /{\cal{R}}
\ee
where $V$ is a typical values of the electric potential produced by the central engine and  ${\cal{R} }\approx 1/ c$ is  the impedance of free space. 

For rotating objects, like \NSs\ and \BHs, this translates to 
\be
L_{EM} \approx \Phi^2 \Omega^2 /c
\label{11}
\ee
where $\Phi$ is the open magnetic flux; for a \NS\ $ \Phi_{NS} \approx B_{NS}  R_{NS}^2 (R_{NS} \Omega_{NS}/c) $, while for a \BH\ (which has no closed field lines)
\be 
\Phi _{BH}\approx B_{BH} R_{BH}^2 
\label{12}
\ee 
and
\be
 \Omega_{BH} \approx a {c \over  R_{BH}}
 \label{13}
 \ee
is the angular velocity of the \BH, $a$ is the Kerr parameter \citep[for  detailed discussion see][]{2011MNRAS.418L..79T}.

Similarly, linear motion of a \Sc\ \BH\ in external \Bf\ with dimensionless velocity $\beta$ produces luminosity \citep[in this case $V\approx \beta B R_{BH}$,][]{2011PhRvD..83f4001L}
\be
L_{EM} = ( GM )^2 B_0^2 \beta^2 /c^3
\label{14}
\ee

One of  the major problem in applying the above relations to produce  \EM\ signal from merging \BHs\ is that, generally,  presence of plasma is required to anchor the \Bf\ (the possibility  that isolated \BHs\ can keep \Bf\ is an exception to this statement, see below).  But 
 very little plasma is expected to be present  in the vicinity of \BHs\ at the moment of the merger: like a kitchen  blender the \BHs\ clear  of matter  the inner few hundreds of  \Sc\ radii   before the merger \citep{2005ApJ...622L..93M,2011PhRvD..84b4024F}. The magnetic field inside the cavity can still remain, created by the currents in the far-away accretion disk, but since at large distances the plasma densities are smaller, the expected \Bf\ is also relatively small so that a weak \EM\ signal is expected  even for the merger of supermassive \BHs\ \citep{2011PhRvD..83l4035L}.  Thus,  little emission is expected instantaneous with the merger. Another possibility is that the recoil from  the merger of   \BHs\ of unequal  masses sends the final \BH\ slamming into the surrounding disk  with a speed of few hundred to few thousand kilometers per second.  This mechanism is expected to produce sub-Eddington luminosities  \citep[\eg][]{2008ApJ...676L...5L}, too little to account for the observed signal.

One of the possible caveat (to the requirement of high circum-merger plasma density to contain the  \Bf) is the 
suggestion by \cite{2011PhRvD..84h4019L,2013ApJ...768...63L} that \BHs\ can keep the magnetic flux for times much longer than predicted by the no hair theorem. (This is due to the fact that the presence of highly conducting plasma around rotating \BHs\ introduces a topological constraint that prohibits \Bfs\ from sliding off the horizon - the no hair theorem \citep[\eg][]{1972PhRvD...5.2439P,MTW} assumes that outside medium is vacuum.)  Though the magnetic retention time is hard to calculate, it is  unlikely that \Bf\ can be kept on the \BHs\ for cosmologically long times during the inspiral. 

\section{Application to possible \EM\ counterpart of  GW150914} 
The outflows powered by the rotational of the central source via the  Blandford-Znajek type mechanism require the \Bfs\  on the \BH\ of the order, Eqns. (\ref{11}-\ref{13}), 
\be
B_{BH} \approx {c^{3/2}\sqrt{ L _{EM} }\over  a GM } = 3 \times 10^{12} {\rm G}
\label{Brot}
\ee
for the inferred parameters of  \BH\ of $M\approx 60 M_\odot$ and the  Kerr parameter $a \approx 0.7$ \citep{2016PhRvL.116f1102A}.

Similarly, the linear motion of  \BHs\ with the Keplerian velocity produces maximal \EM\ power, see Eq. (\ref{14}) with $\beta \approx 1$, of
\be
L_{EM, max} \approx { (G M B_0)^2 \over c^3} 
\ee 
and  requires
\be
B_0 \approx {c^{3/2} \sqrt{L} \over GM } = 2 \times 10^{12} \, {\rm G}.
\label{Blin}
\ee

Both estimates (\ref{Brot}) and (\ref{Blin}) require exceptionally high magnetic fields, typical of young \NSs\ and {\it  exceeding by many orders of magnitude the \Bfs\ expected from accretion of the interstellar material.}  For example,  regardless of the central mass, and 
parametrizing accretion luminosity  required to contain \Bfs\  (\ref{Brot}-\ref{Blin}) as 
\be
L _{EM}= \eta \dot{M} c^2
\ee
($\eta \sim 0.1$ is some efficiency factor), the required accretion rate is 
$ \dot{M} \approx 5 \times 10^{-5} M_\odot $ sec$^{-1}$ $= 1.5 \times 10^3 M_\odot $ yr$^{-1}$ . This is an unreasonably high accretion rate from the ISM or even in  a binary system. Association of the BH merger with interiors of massive stars, that could possibly provide the required accretion rates \citep[\eg in a double collapsar-type scenario akin to][]{1999ApJ...524..262M}, see  \cite{2016arXiv160204735L}, contradicts the claimed  $\sim 1$ second duration of the signal.

Collimation of the \EM\ outflows, though reducing the overall energetics, is not likely to be important. First, the gravitation waves signal is only slightly anisotropic \citep{2016arXiv160203840T}. Highly anisotropic \EM\  emission would  make the  contemporaneous detection highly unlikely, approximately by the ratio of the anisotropy solid angle over $4\pi$. Also, nozzle-type outflow collimation requires confinement (\Bf\ can somewhat  reduce the required confining pressure below the typical energy density in the jet, but cannot eliminate it completely). It is not expected that the circum-merger medium has sufficient density for confinement. Also note that isolated rotating black holes that retain 
\Bf\  flux would produce equatorially-collimated outflows \citep[][edge-on orientation is slightly disfavored fore GW150914, \cite{2016arXiv160203840T}]{2011PhRvD..84h4019L,2014ApJ...788..186N}.  Linear motion of \BHs\ in external \Bf\ produces a   dual  pair of jets \citep{2010Sci...329..927P,2011PhRvD..83f4001L}.

Another possible subtlety is that the {\BHs}'s orbital motion and/or the spin can amplify the external \Bf\  in a way analogous to the two conducting spheres dynamo \citep{1978mfge.book.....M,1993AIPC..280..992B}. 
This possibility, likely, does not apply to the case of  two  orbiting/rotating \BHs\ -  \BHs\ are, qualitatively, bad conductors \citep{ThornMembrane}, so that the external \Bf\ lines  would slide along the spinning  horizon \citep[see,\eg numerical calculations of][]{2010PhRvD..82d4045P}.

Finally, requiring that the \Bf\ is produced by electric charge on one of the \BHs\ \citep{2016arXiv160204542Z} would imply a charge $Q = {G M \sqrt{L_{EM}}/c^{5/2}} = 5\times 10^{16}$ coulombs. This   horrific  amount of electric charge  would  have  produced the \Ef\ near the horizon  $E =  \sqrt{L_{EM}} c^{3/2} /( GM ) = 2 \times 10^{12}$ in cgs units (statvolts per centimeter), amounting  to nearly  $5\% $ of the quantum Schwinger  field $E_Q = m_e^2 c^3 /(\hbar e)$.

\section{Conclusion}

We discuss the physical requirements that the possible observation of the \EM\ signal contemporaneous with GW150914 imposes on the circum-merger environment.  We find that the required physical parameters at the source exceed by many orders of magnitude what is expected in realistic astrophysical scenarios. Given  these constraints, and that  the chance probability of the signal  is not particularly  low, and no signal was detected by other satellites,  we conclude that  the Fermi GBM signal contemporaneous with  GW150914  is unrelated to the BH merger.

I would like to thank Rodolfo Barniol Duran,  Roger Blandford,  Patrick Brady, Dimitrios Giannios, Maria  Petropoulou and Ksenia Ptitsyna for discussions and comments on the manuscript.  
\bibliographystyle{apj}
%\bibliography{~/Home/PulsarRadio/PulsarBib}
 \bibliography{/Users/maxim/Home/Research/BibTex}

\end {document}